\documentstyle[multicol,aps,epsf]{revtex}

\begin{document}
\title{Aging and its Distribution in Coarsening Processes}
\author{L.~Frachebourg\footnote{Present address: Laboratoire de
Physique Statistique, ENS, 24 rue Lhomond, 75231 Paris Cedex 05,
FRANCE}, P.~L.~Krapivsky, and S.~Redner} 
\address{Center for Polymer
Studies and Department of Physics, Boston University, Boston, MA 02215}
\maketitle
\begin{abstract}

We investigate the age distribution function $P(\tau,t)$ in prototypical
one-dimensional coarsening processes.  Here $P(\tau,t)$ is the
probability density that in a time interval $(0,t)$ a given site was
last crossed by an interface in the coarsening process at time $\tau$.
We determine $P(\tau,t)$ analytically for two cases, the (deterministic)
two-velocity ballistic annihilation process, and the (stochastic)
infinite-state Potts model with zero temperature Glauber dynamics.
Surprisingly, we find that in the scaling limit, $P(\tau,t)$ is
identical for these two models.  We also show that the average age, {\it
i.\ e.}, the average time since a site was last visited by an interface,
grows linearly with the observation time $t$.  This latter property is
also found in the one-dimensional Ising model with zero temperature
Glauber dynamics. We also discuss briefly the age distribution in
dimension $d\geq 2$.

\medskip
{PACS numbers:  02.50.Ga, 05.70.Ln, 05.40.+j}
\end{abstract}
\begin{multicols}{2}

\section{INTRODUCTION AND PROBLEM STATEMENT}

Coarsening underlies various natural non-equilibrium processes, {\it e.\
g.}, phase separation in binary alloys, grain growth, and growth of soap
bubbles\cite{langer}.  A common feature of coarsening phenomena is the
scale-invariant morphology that arises in the late
stage\cite{langer,bray}.  Such a behavior is a signature of dynamical
scaling.  If dynamical scaling holds, the average domain size,
$\ell(t)$, typically exhibits algebraic growth, $\ell(t)\sim t^{1/z}$.

It has recently been appreciated that knowledge of the dynamical
exponent $z$ does {\it not} provide a comprehensive description of the
coarsening dynamics.  In particular, the exponent $\lambda$ which
describes the dependence of the autocorrelation function $A(t)\equiv
\langle s(x,0)s(x,t)\rangle$, where $s(x,t)$ is the order parameter at
position $x$ and time $t$, on the average domain size,
$A(t)\sim\ell(t)^{-\lambda}$\cite{bray,fisher,desai}, and the exponent
$\theta$ which characterizes (in magnetic language) the fraction of
spins which have never flipped, $P_0(t)\sim
t^{-\theta}$\cite{dbg,kbr,bdg}, were found to be independent of the
dynamical exponent $z$.  The latter quantity, $P_0(t)$, naturally
suggests the generalization to $P_n(t)$, the fraction of spins which
have flipped exactly $n$ times up to time $t$\cite{elp} as a detailed
and fundamental characterization of the temporal history of spin flips.

In this study, we investigate a related aspect of this temporal history
by focusing on the time $\tau$ when the last spin flip occurs (Fig.~1).
More generally, we may introduce $P_n(\tau,t)$ as the probability that a
given spin flips $n$ times up to time $t$ {\it and\/} that the last spin
flip occurs at time $\tau$.  Here we investigate $P_+(\tau,t)$ which
focuses on the last spin flip and does not specify the total number of
flips, $P_+(\tau,t)=\sum_{n\geq 1}P_n(\tau,t)$.  If we view a spin as
being ``reborn'' each time it flips, then $P_+(\tau,t)$ gives the density
of spins of ``age'' $t-\tau$.  There is also a finite fraction of spins
which never flipped yet; these spins should be treated as spins of age $t$.
The total age distribution density of the spins is therefore

\begin{equation}
P(\tau,t)=P_0(t)\delta(\tau)+P_+(\tau,t).
\label{total}
\end{equation}
The density $P(\tau,t)$ should satisfy the normalization condition
$\int_0^t d\tau P(\tau,t)=1$, while the average age of the system is
defined via

\begin{eqnarray}
T &=&\int_0^t d\tau (t-\tau)P(\tau,t)\nonumber\\
  &=&tP_0(t)+\int_0^t d\tau (t-\tau)P_+(\tau,t). 
\label{age}
\end{eqnarray}

\begin{figure}
\narrowtext
\epsfxsize=\hsize
\epsfbox{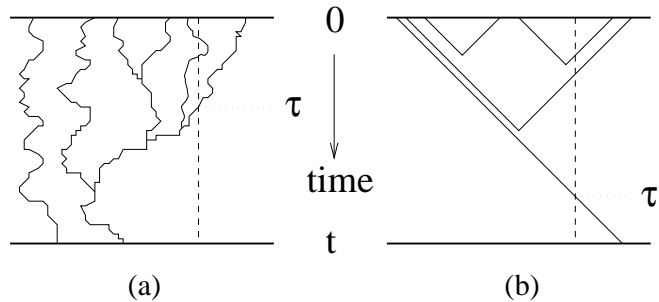}
\vskip 0.2in
\caption{
Graphical definition of $P(\tau,t)$ for one-dimensional
coarsening processes.  At the point marked by the dashed line, the spin
last flips, or equivalently, is visited by a domain wall, at time
$\tau$.  The specific examples shown are: (a) the infinite-state Potts
model (in which the domain walls undergo diffusive single-species
coalescence) and (b), the deterministic coarsening of a 3-state system
with cyclic interactions (in which the domain walls undergo ballistic
single-species annihilation).
\label{fig1}}
\end{figure}

The age distribution $P(\tau,t)$ will be of primarily importance in
systems with history-dependent dynamics, such as glassy
systems\cite{angell}, and in systems with infinite memory where actual
aging takes place.  Generally, when a two-time correlation function
${\cal C}(\tau,t)=\langle s(x,\tau)s(x,t)\rangle$ becomes a function of
a single variable $\tau/t$, instead of being a function of $t-\tau$ (as
in an equilibrium system), this is interpreted as a signature of
aging\cite{bou1,bou2,bou3}.  According to this definition, aging is a
characteristic of coarsening processes and the scaling dependence
$P(\tau,t)\simeq t^{-1}f(\tau/t)$ has been found in a number of
pertinent examples\cite{kolja,slava}.  

The age distribution will also play a fundamental role when the dynamics
of a system is explicitly time dependent.  A potentially interesting
situation is that of the ``adaptive'' voter model.  The conventional
voter model\cite{lig} is a two state lattice system in which a voter
(site) randomly chooses one of its nearest neighbors and assumes the
state of this neighbor.  In the adaptive extension of this model the
probability that a given voter changes its opinion depends on the local
environment (as in the usual voter model) {\it and\/} on the time
interval since this particular voter last changed its opinion.  This
might be viewed as a model to describe the increasing conservatism of
people when they are not stimulated by contact with those of differing
opinions.  This adaptive voter model exhibits rather unexpected
coarsening dynamics which is ultimately driven by the underlying age
distribution \cite{adaptive}.  In particular, we find coarsening for all
spatial dimensions, while the conventional voter model coarsens only for
spatial dimension $d\leq 2$.

In the following two sections, we consider the age distribution for two
specific one-dimensional coarsening processes for which exact results
can be obtained.  In Sec.\ II, we first treat a deterministic 3-state
model of coarsening in which the dynamics of the domain walls is simply
that of two-velocity ballistic annihilation.  Because of this
equivalence, it is possible to obtain the exact expression for
$P(\tau,t)$ by simple means.  In Sec.\ III, we investigate the age
distribution in two stochastic coarsening models.  The first is the
infinite-state Potts model in which the domain wall dynamics is simply
diffusion-limited coalescence process, which may be represented as
$A+A\to A$.  We find that the scaling form of the age distribution is
identical to that found in the deterministic coarsening process.  We
also consider the age distribution for the Ising model with
zero-temperature Glauber dynamics in which the domain wall dynamics
coincides with single-species diffusion-limited annihilation process,
which may be represented as $A+A\to 0$.  In this case, the age
distribution has a bimodal ``smiling'' form as a function of $\tau$, a
result which can be understood intuitively.  We then discuss the age
distribution for the dynamical Ising model in dimension $d\geq 2$ and
give an exact expression for the distribution in the mean-field limit.
Sec.\ IV gives a brief summary and outlook.

\section{AGING IN A DETERMINISTIC MODEL OF COARSENING}

We first examine the age distribution in a deterministic coarsening
model which describes phase ordering dynamics in a cyclic
one-dimensional system with three equilibrium states, $A$, $B$, and $C$.
The dynamics is cyclic so that the $B$ phase invades the $A$ phase, $C$
invades $B$, and $A$ invades $C$.  Corresponding to this dynamics,
interfaces between dissimilar domains move toward the subordinate domain
with a fixed velocity.  A domain which is besieged by two dominant
domains shrinks and eventually disappears, leading to the merging of the
neighboring domains.  The interfaces therefore undergo ballistic motion
with annihilation occurring whenever two interfaces meet.  These rules
are precisely those of the ballistically-driven single-species
annihilation reaction.  The simplicity and rich phenomenology of this
reaction has stimulated extensive fundamental
work\cite{els,brl,krl,Piasecki,Droz}, as well as related applications to
growth processes\cite{Krug,sekimoto,ben,gold}, and the dynamics of
interacting populations\cite{bramson,fisch,lpe}.

We start by describing the behavior\cite{els,Piasecki} of the ballistic
annihilation model for the domain walls.  In this model, the density of
right-moving and left-moving walls is equal, with velocities which can
be taken to be $\pm 1$ without loss of generality.  From the exact
solution\cite{els}, the probability $S(t)$ for an arbitrary
interface to survive up to time $t$ is

\begin{equation}    
S(t)=e^{-2t}[I_0(2t)+I_1(2t)].
\label{st}
\end{equation}
Here $I_j$ denotes the modified Bessel function of order $j$, the
initial spatial distribution of interfaces is assumed to be Poissonian
(no correlations), with the initial densities of $\pm$ interfaces taken
to be equal 1/2.

To obtain the age distribution for the coarsening process induced
by this domain wall dynamics, first consider $P_0(t)$, the fraction
of space that has not been crossed by any interface in the time interval
$(0,t)$.  One can interpret $P_0(t)$ as the probability that a
stationary ``target'' particle, which is placed at the origin, for
example, is not hit by any moving domain wall.  It is convenient to
consider an auxiliary one-sided problem with interfaces distributed only
to the right of the origin.  For this case, the survival probability of
the stationary particle, $S_0(t)$, is\cite{krl}

\begin{equation}    
S_0(t)=e^{-t}[I_0(t)+I_1(t)].
\label{s0t}
\end{equation}
Indeed, the relative velocity between a stationary particle and its
reaction partner is a factor of two smaller than the relative velocity
between two moving reaction partners.  Hence, $S_0(t)=S(t/2)$, and
Eq.~(\ref{s0t}) follows from Eq.~(\ref{st}).  Clearly, the survival
probability $P_0(t)$ in the original two-sided problem is

\begin{equation}    
P_0(t)=S_0(t)^2.  
\label{p0t}
\end{equation}

The continuous part of the age distribution $P_+(\tau,t)$ can also be
expressed in terms of the survival probabilities $S(t)$ and $S_0(t)$.
We first note that for the origin to be crossed by an interface during
the time interval $(\tau,\tau+d\tau)$, a left moving interface should be
initially located in the spatial interval $\tau<x<\tau+d\tau$, or a
right moving interface should be located in the spatial interval
$-\tau-d\tau<x<-\tau$.  Each of these events occurs with probability
$d\tau/2$ for an initial interface density of unity.  

Suppose that the origin is crossed by a left moving interface (Fig.~2).
Then this interface will ultimately be annihilated with some right
moving interface at some future time $t_1$, which satisfies $t_1>\tau$.
If $t_1>(t+\tau)/2$, then the origin cannot be crossed by a right
moving interface during the time interval $(\tau,t)$. The contribution
of these type of configurations to $P_+(\tau,t)$ is

\begin{equation}    
S\left({t+\tau\over 2}\right)S_0(t-\tau).  
\label{conf1}
\end{equation}
The first factor is just the probability that the left moving
interface survives up to time $(t+\tau)/2$.  The latter factor in
Eq.~(\ref{conf1}) is the probability that the initial location of the
left moving interface has not been crossed by any other left moving
interface during the time interval $(0,t-\tau)$ which, in turn, ensures
that the origin remains uncrossed from the right during the time
interval $(\tau,t)$.

\begin{figure}
\narrowtext
\epsfxsize=\hsize
\epsfbox{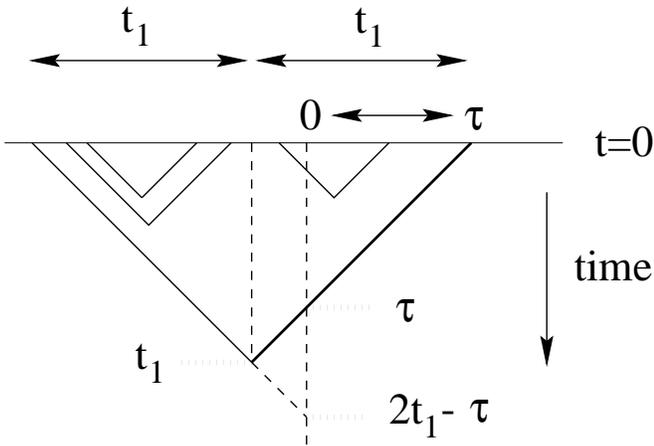}
\vskip 0.2in
\caption{
Illustration of a typical configuration which contributes
to $P_+(\tau,t)$ in the deterministic coarsening process generated by
domain walls which undergo ballistic single-species annihilation.  The
left-moving domain wall trajectory which crosses the origin at time
$\tau$ is shown as a heavy line.  This domain wall is annihilated at a
time $t_1>\tau $ such that any right-moving trajectory cannot reach the
origin before time $t=2t_1-\tau$.
\label{fig2}}
\end{figure}

Consider now the complementary situation when the left-moving interface
which crosses the origin in the time interval $(\tau,\tau+d\tau)$
survives to time $t_1$ with $\tau<t_1<(t+\tau)/2$.  In this case,
additional right-moving interfaces can cross the origin before time $t$.
The contribution of such configurations to $P_+(\tau,t)$ is

\begin{equation}    
S_0(t-\tau)\int_{\tau}^{t+\tau\over 2} S_0(t-2t_1+\tau)[-\dot S(t_1)]dt_1.
\label{conf2}
\end{equation}
Here $S_0(t-\tau)$ again ensures that the origin remains uncrossed from
the right during the time interval $(\tau,t)$.  Similarly,
$S_0(t-2t_1+\tau)$ guarantees that the origin remains uncrossed from the
left.  Finally, $-\dot S(t_1)dt_1$ is the probability that the left
moving interface is annihilated in the time interval $(t_1,t_1+dt_1)$.
Combining these contributions, gives the final exact expression for the
age distribution density

\begin{eqnarray}    
P(\tau,t)&=&S_0(t)^2\delta(\tau)
+S\left({t+\tau\over 2}\right)S_0(t-\tau)\nonumber\\
&-&S_0(t-\tau)\int_{\tau}^{t+\tau\over 2} S_0(t-2t_1+\tau)\dot S(t_1)dt_1.
\label{main}
\end{eqnarray}

The singular part of the age distribution, $S_0(t)^2\delta(\tau)$,
corresponds to the fraction of space that has not been traversed by any
interface; in the long-time limit, this fraction decays as $t^{-1}$.  To
determine the asymptotic behavior of the continuous part of the age
distribution, we substitute into Eq.~(\ref{main}) the asymptotic
expressions, $S(t)\sim 1/\sqrt{\pi t}$ and $S_0(t)\sim \sqrt{2/\pi t}$,
which are found by using the asymptotic relations for the modified
Bessel functions, $I_j(z)\to e^z/\sqrt{2\pi z}$ as $z\to
\infty$ and $j$ fixed\cite{bender}.  The contribution of the
third term of Eq.~(\ref{main}) turns out to be asymptotically
negligible, while the second term leads to the scaling form,

\begin{equation}    
P_+(\tau,t)\simeq t^{-1}f(\xi),
\label{scal}
\end{equation}
in the scaling limit 
\begin{equation}  
t\to\infty,    \quad
\tau\to\infty, \quad
\xi=\tau/t,
\label{scalv}
\end{equation}
with the scaling function given by
\begin{equation}    
f(\xi)={2\over\pi}\,{1\over \sqrt{1-\xi^2}}.
\label{scalf}
\end{equation}

A prominent feature of the age distribution is that $\tau$ scales as
$t$.  That is, the average age,

\begin{equation}
T=\langle t-\tau\rangle\simeq 
t\int_0^1 d\xi\,(1-\xi)f(\xi)\simeq \left(1-{2\over\pi}\right)t,
\label{time}
\end{equation}
grows linearly with the observation time $t$.

\section{AGING IN STOCHASTIC MODELS OF COARSENING}

The ballistic annihilation model is perhaps the simplest one-dimensional
coarsening process with {\it deterministic} dynamics.  We now consider
simple examples of one-dimensional coarsening processes with {\it
stochastic} dynamics.  Consider first the $q$-state Potts model for
$q=\infty$, with zero temperature Glauber dynamics and with the initial
condition where each spin is in a different state.  The dynamics
proceeds as follows: during the time interval $dt$ a given spin assumes
the state of one of its nearest neighbor with overall probability
$dt/2$.  In one dimension, the interfaces between domains of identical
spins therefore diffuse and coalesce whenever two domains meet.  The
domain wall dynamics is thus identical to the diffusion-limited
coalescence reaction, which may be represented as $A+A\rightarrow A$.

Because of this equivalence between the Potts model and the coalescence
reaction, the age distribution can be calculated exactly.  Since
interfaces coalesce upon colliding, only the interfaces which are the
nearest neighbors of a particular site are important in determining its
age distribution.  In constructing the age distribution, first note that
the spin will not change its color up to time $t$ if neither of the two
neighboring interfaces reaches the spin.  The probability $P_0(t)$ is
thus equivalent to the square of the probability $Q(t,1)$ that a random
walker on a lattice starting at position $x_0=1$ will not reach the
origin up to time $t$. The probability $Q(t,1)$ is readily
computable\cite{glauber} and gives the fraction of ``persistent'' spins:

\begin{equation}
P_0(t)=\left(e^{-t}[I_0(t)+I_1(t)]\right)^2.
\label{p0potts}
\end{equation}

To compute the contribution to the age distribution from configurations
where an interface has previously reached the spin (which we may take to
be at the origin), let us assume that this spin takes on a new color
from its left neighbor at time $\tau$.  This spin is now the right
extremity of a domain of same color spins (see Fig.~3).

\begin{figure}
\narrowtext
\epsfxsize=\hsize
\epsfbox{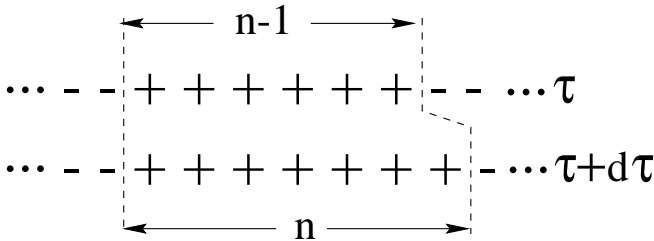}
\vskip 0.2in
\caption{
Illustration of one process which enters in the
computation of $P_+(\tau,t)$ for the infinite-state Potts model.  Shown
is the spin configuration at times $\tau$ and $\tau+d\tau$ just as one
spin changes its state.  For the state this spin to remain unchanged
until time $t$, both the domain wall a distance 1 to the right and the
domain wall a distance $n$ to the left must not reach the position of
the newly-flipped spin.
\label{fig3}}
\end{figure}

Let the size of this domain be $n$.  The position of the interface
which defines the left edge of this domain is distributed according to
the domain size distribution $F(n-1,\tau)$.  The spin at the origin will
then not change its color up to time $t$ if the two surrounding
interfaces do not cross the origin.  The continuous part of the age
distribution can thus be written as
 
\begin{equation}
P_+(\tau,t)=\sum_{n=2}^{\infty}F(n-1,\tau)Q(t-\tau,n)Q(t-\tau,1).
\label{p+potts}
\end{equation}
The last factor is just the probability that the domain which is one
lattice spacing to the right of the spin at the origin does not reach
the origin between time $\tau$ and time $t$, while the first two factors
given the corresponding probability for the left-neighboring domain
which is a distance $n$ from the origin.

Each of the factors in this equation are well known.  The domain size
distribution is given by $F(n-1,\tau)= E(n-1,\tau)- 2E(n,\tau)+
E(n+1,\tau)$, where $E(k,t)$ is the probability to find at least $k$
successive spins of the same color at time $t$\cite{ben-avraham}.  For a
discrete lattice system, this latter distribution satisfies a lattice
diffusion equation, with boundary condition $E(0,t)=1$ and initial
condition $E(k,0)=\delta_{k,0}$, corresponding to the initial condition
where each spin is different.  The expression for $E(k,t)$
is\cite{glauber}

\begin{equation}
E(k,t)=1-e^{-2t}\left[I_0(2t)+2\sum_{j=1}^{k-1}I_j(2t)+I_k(2t)\right]
\label{ekt}
\end{equation}
and thus
\begin{equation}
F(n-1,\tau)={e^{-2\tau}\over\tau}\,nI_n(2\tau). 
\label{fkt}
\end{equation}
In a similar vein, the probability $Q(t,k)$ that a random walker
which starts at $x=k$ does not hit the origin during the time interval
$(0,t)$ is\cite{glauber}
\begin{equation}
Q(t,k)=e^{-t}\left[I_0(t)+2\sum_{j=1}^{k-1}I_j(t)+I_k(t)\right].
\label{qkt}
\end{equation}
So we finally obtain
\begin{eqnarray}
P_+(\tau,t)={e^{-2t}\over\tau}\,\left[I_0(t-\tau)+I_1(t-\tau)\right]
\,\sum_{n=1}^\infty nI_n(\tau)\nonumber\\
\times\left[I_0(t-\tau)+2\sum_{k=1}^{n-1}I_k(t-\tau)+I_{n}(t-\tau)\right].
\label{ptpottssol}
\end{eqnarray}

In the scaling limit (\ref{scalv}), the dominant contribution to the sum
in Eq.~(\ref{ptpottssol}) is provided by terms with $n\propto\sqrt{t}$.
In this region we use the asymptotic form of the Bessel functions
$I_n(t)\simeq \exp(t-n^2/2t)/\sqrt{2\pi t}$.  A lengthy but elementary
computation then yields

\begin{equation}
P_+(\tau,t)\simeq {2\over \pi \sqrt{t^2-\tau^2}}
\label{scalptpotts}
\end{equation}
which is exactly of the same form as Eqs.~(\ref{scalv})--(\ref{scalf}).
At first sight, it may seem surprising to find the same scaling
function, as well as the same expression for $P_0(t)$, as in the
ballistic annihilation problem.  Indeed, Eq.~(\ref{s0t}) can be computed
from a mapping of the initial distribution of the interfaces onto a
random walk process.  $S_0(t)$ can then be computed in the same way
as the probability $Q(t,1)$ shown above.  Whenever we can determine a
property of the infinite-state Potts model via the behavior of two
independent random walks, we should recover the same results as in the
ballistic annihilation problem.  Nevertheless, some properties of these
two systems are very different.  For example, the domain size
distribution in ballistic annihilation exhibits a non-trivial behavior
which is characterized by an infinite number of singularities
\cite{Piasecki,laupaul}.

Let us now consider the age distribution of spins in the 2-state Potts
model with zero temperature spin-flip dynamics, {\it i.\ e.}, the
kinetic Ising-Glauber model\cite{glauber}.  Since the solution for
$P_0(t)$ in the Ising-Glauber model is difficult\cite{der}, one can
anticipate that calculation of $P_+(\tau,t)$ is also subtle.  We
therefore study this problem numerically and give heuristic arguments to
explain the limiting behaviors of the age distribution $P_+(\tau,t)$.

Our numerical results, which are based on simulations of the equivalent
$A+A\to 0$ reaction process, confirm that the scaling ansatz
(\ref{scal})--(\ref{scalv}) still applies (Fig.~4).  

\begin{figure}
\narrowtext
\epsfxsize=\hsize
\epsfbox{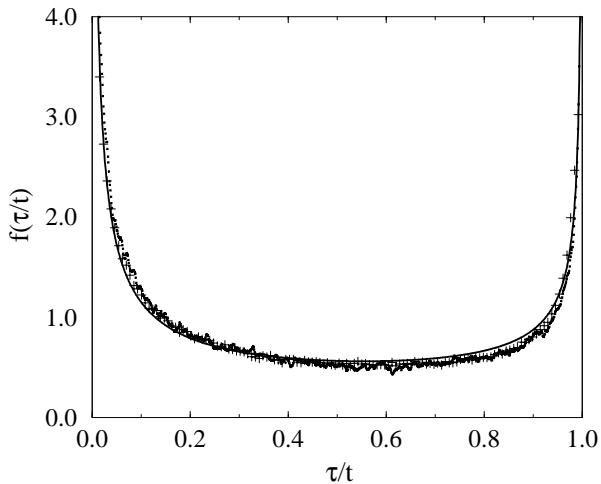}
\caption{Simulation data for the age distribution in the one-dimensional
Ising-Glauber model.  Shown is the scaling function $f(\xi)$ versus
$\xi$ for $t=(1.5)^{12}$ $(+)$ and $t=(1.5)^{17}$ $(\circ)$, with the
latter data averaged (smoothed) over 5 consecutive points.  The solid
line is the guess $f_{\rm guess}(\xi) ={\cal B}\,
\xi^{-5/8}(1-\xi)^{-1/2}$, with ${\cal B}= 0.259349\ldots$ as explained
in the text.
\label{fig4}}
\end{figure}

The singular behavior of the scaling part of the age distribution
function $f(\xi)$ in the limits $\xi\downarrow 0$ and $\xi\uparrow 1$
can be accounted for by matching to the known behaviors in these limits.
When $\tau={\cal O}(1)$, $P_+(\tau,t)\sim P_0(t)\sim
t^{-3/8}$\cite{der}.  Matching this with Eq.~(\ref{scal}) at
$\xi=\tau/t={\cal O}(t^{-1})$ implies the $f(\xi)\sim \xi^{-5/8}$ as
$\xi\downarrow 0$.  This asymptotic behavior agrees well with our
simulations.  In the opposite limit of $\tau\to t$, the corresponding
limiting form of the age distribution is determined by domain walls
which have crossed the origin at time $\tau$ close to $t$ --- this
happens with probability $t^{-1/2}$, since the number of domain walls
decreases with time as $t^{-1/2}$\cite{glauber}.  The diffusing domain
wall should then not cross the origin again in the following time
interval $(\tau,t)$ --- this happens with probability
$(t-\tau)^{-1/2}$\cite{feller}.  Thus, $P(\tau,t)\sim
t^{-1/2}(t-\tau)^{-1/2}$, which implies that $f(\xi)\sim (1-\xi)^{-1/2}$
as $\xi\uparrow 1$, in agreement with our numerical results.  Indeed,
the product of these two asymptotic forms, $f_{\rm guess}(\xi)={\cal
B}\, \xi^{-5/8}(1-\xi)^{-1/2}$ provides a reasonable fit to the data
over most of the range of $\xi$.  If one uses this guess over the entire
range of $\xi$, then the normalization condition $\int d\xi f_{\rm
guess}(\xi)=1$, requires the numerical prefactor to be ${\cal
B}={\Gamma(7/8)\over \Gamma(3/8)\Gamma(1/2)} = 0.259349\ldots$.

For the general $q$-state Potts model with zero-temperature Glauber
dynamics, we may also expect that the age distribution scales, with the
limiting behaviors of the scaling function given by

\begin{equation}
f(\xi)\sim \cases{\xi^{\theta(q)-1} &   $\xi\downarrow 0$,\cr
                         &\cr
                        (1-\xi)^{-1/2}     &   $\xi\uparrow 1$.  \cr}
\label{limits}
\end{equation}
The persistence exponent $\theta(q)$, found analytically in
Ref.~\cite{der}, increases from 3/8 to 1 as $q$ increases from 2 to
$\infty$.  Thus the ``smiling'' form of the age distribution in the
Ising case (Fig.~4) gradually transforms into the half-smiling form of
the infinite-state model (see Eq.~(\ref{scalptpotts})).

In more than one dimension, aging of spins in the kinetic Ising model is
expected to depend on the temperature.  If an initially disordered
system is quenched to a final temperature $T_f>0$, the average age is
expected to be finite for all $d>1$.  This follows because for
non-conserved dynamics, even spins embedded within a large region of
aligned spins will flip at a finite rate for all positive temperatures.
On the other hand, for a quench to zero temperature, we anticipate that
the average age will grow with time, since spin flips can occur only at
interfaces, and these eventually disappear.  To test this expectation,
we performed numerical simulations of the two-dimensional kinetic
Ising-Glauber model on the square lattice and found that the average age
of the spins grows linearly in time and that scaling still applies.
Moreover, the age distribution function has the same qualitative
``smiling'' form of the one-dimensional system (Fig.~4).

In the small-age limit, $t-\tau\ll t$, the numerical data suggests a
behavior of the age distribution which is consistent with $P(\tau,t)\sim
t^{-1/2}(t-\tau)^{-1/2}$.  To understand this result, which is identical
to that of the one-dimensional counterpart, first note that the density
of domain walls decays as $t^{-1/2}$.  This arises because for
non-conserved dynamics, the average domain size grows as
$t^{1/2}$\cite{bray} and domains appear to be compact.  Consequently,
the domain wall density is expected to be the reciprocal of the average
domain size.  The perimeter of a domain has typically a vicinal shape,
with the kinks and antikinks which define terraces undergoing diffusive
motion (this diffusion does not cost energy and is therefore allowed at
zero temperature).  This diffusional motion is one-dimensional in
character and thus a step (either kink or antikink) which has crossed a
bond at time $\tau$ will not cross it again in the following time
interval $(\tau,t)$ with probability $(t-\tau)^{-1/2}$.  The age
distribution is then given by the product of step density and the above
no return probability, which gives $P(\tau,t)\sim
t^{-1/2}(t-\tau)^{-1/2}$.  In fact, the evolution of interfaces is much
more involved process -- kinks and antikinks annihilate upon colliding,
spin-flips at the corner give birth to a pair of steps (horizontal and
vertical) -- but in the small-age limit these additional complexities
should not qualitatively affect the age distribution.

In the large-age limit, $\tau\ll t$, the scaled age distribution is
expected to behave as $f(\xi)\sim \xi^{\theta-1}$, similarly to one
dimension.  Indeed, we confirmed numerically such power-law behavior and
found that $\theta\approx 0.21$ provides the best fit to our data.  This
is consistent with previous simulations of the two-dimensional
Ising-Glauber model for which the fraction of persistent spins,
$P_0(t)$, was found to decay as $t^{-0.22}$\cite{dbg,stauf}.

To determine the form of the age distribution for the kinetic
Ising-Glauber model in higher dimensions, we apply a mean-field
approach.  It is simple to solve for $P(\tau,t)$ in the mean-field limit
({\it e.\ g.}, for the Ising model on a complete graph) since the
dynamics in the zero-temperature case is simple: Spins from the majority
phase do not change their state, while spins from the minority phase
change their state with a constant rate which we may set equal to one.
Suppose that the system starts from an initial condition where the
fraction of $+$ and $-$ spins is equal to $p$ and $q=1-p$, respectively
(with $p\geq q$ without loss of generality).  Clearly, the fraction of
spins which never change their state until time $t$ is equal to
$p+qe^{-t}$.  The probability that a minority spin changes its state in
the time interval $(\tau,\tau+d\tau)$ is equal to $e^{-\tau}d\tau$.
Thus,

\begin{equation}
P(\tau,t)=\left(p+qe^{-t}\right)\delta(\tau)+qe^{-\tau}.
\label{mfising}
\end{equation}
This result violates the scaling form of Eq.~(\ref{scal}) but still
implies that the average age (see Eq.~(\ref{age})) increases linearly in
time:

\begin{eqnarray}
T &=&\left(p+qe^{-t}\right)t+q\left(t-1+e^{-t}\right)\nonumber\\
  &=& t-q\left(1-e^{-t}-te^{-t}\right).
\label{mfage}
\end{eqnarray}

\section{SUMMARY AND OUTLOOK}

The age distribution in one-dimensional coarsening processes has been
investigated by analytical and numerical techniques.  These approaches
indicate that the average age grows linearly with the observation time
of the system.  Exact results for two prototypical coarsening processes,
the deterministic ballistic annihilation and the stochastic
infinite-state Potts model with zero temperature Glauber dynamics have
been obtained.  For the general $q$-state Potts model with zero
temperature Glauber dynamics, asymptotic behaviors have been
established.  

Various results for the aging of spins in the Ising-Glauber model in
general dimension have been obtained.  The interesting situation, for
non-conserved spin-flip dynamics, is that of zero temperature where
domain walls ultimately disappear so that the system undergoes aging.
In particular, numerical results in two dimensions were found to be
qualitatively similar to corresponding one-dimensional results.  We
anticipate that the bimodal ``smiling'' form of the age distribution
will arise for all spatial dimension $d<4$.  When $d\geq 4$, however,
the age distribution is expected to exhibit features similar to the
easily-derived mean-field solution (see Eq.~(\ref{mfising})).  In
particular, the fraction of spins which never flip should saturate at a
finite value even in the symmetric case of $p=q=1/2$.  This has
apparently been observed \cite{stauf}, although it is hard to
definitively settle this issue by numerical means, especially in the
marginal case of $d=4$.

It is worth noting that for the models discussed in this work, the only
possibilities found are systems where the average age saturates to a
finite value or where the average age increases linearly in time.  The
saturation of the age in first class of systems arises because a steady
state is reached.  On the other hand, for systems which coarsen is is
perhaps worth investigating whether there are examples where the average
age grows slower than linear in time.  Numerical evidence shows that the
average age in the two-dimensional voter model is growing slower than
linearly and perhaps logarithmically in time.  This intriguing
possibility merits further consideration.

For the coarsening processes examined in this work, the dynamics
determines the age distribution.  It may be instructive to study models
with feedback, in which the aging process influences the coarsening
dynamics\cite{adaptive}.  The adaptive voter model is one such example.
Another possibly intriguing extension would be to consider coarsening
processes with conservative dynamics.

\vskip 0.16in 
The research reported here was supported in part by the Swiss National 
Foundation, by the ARO (grant DAAH04-93-G-0021), and by the
NSF (grant DMR-9219845).  We also thank J. F. Mendes for helpful
discussions.

\end{multicols} 
\end{document}